\documentclass[10pt,letterpaper]{article}
\usepackage[top=0.85in,left=2.75in,footskip=0.75in]{geometry}
\usepackage{algorithm}
\usepackage{algpseudocode}
\usepackage[dvipsnames]{xcolor}
\usepackage{amsmath,amssymb}
\usepackage{graphicx}
\usepackage{subcaption}
\usepackage{caption}

\usepackage[percent]{overpic} 

\usepackage{changepage}
\usepackage{float}
\usepackage{textcomp,marvosym}

\usepackage{cite}
\usepackage{booktabs}
\usepackage{nameref,hyperref}

\usepackage[right]{lineno}

\usepackage[expansion=false]{microtype}

\DisableLigatures[f]{encoding = *, family = * }

\usepackage[table]{xcolor}

\usepackage{array}

\newcolumntype{+}{!{\vrule width 2pt}}

\newlength\savedwidth

\usepackage[aboveskip=1pt,labelfont=bf,labelsep=period,justification=justified,singlelinecheck=off]{caption}

\makeatletter
\renewcommand{\@biblabel}[1]{\quad#1.}
\makeatother

\usepackage{lastpage,fancyhdr,graphicx}
\usepackage{epstopdf}
\fancyheadoffset[L]{2.25in}
\fancyfootoffset[L]{2.25in}
\begin{document}
\vspace*{0.2in}

\begin{flushleft}
{\huge
\textbf\newline{Regret, Uncertainty, and Bounded \\ 
Rationality in Norm-Driven Decisions} 
}
\newline
\\
Christos Charalambous\textsuperscript{1*}
\\
\bigskip
\textbf{1} University of Cyprus, Department of Economics, PO Box 20537, 1678 Nicosia, Cyprus
\\
\bigskip

%
%





* charalambous.christos.2@ucy.ac.cy

\end{flushleft}
\section*{Abstract}

This study introduces an agent-based model to study how regret, uncertainty, and social norms interact to shape vaccination behavior during epidemics. The model integrates three behavioral mechanisms—anticipated regret, evolving norms, and uncertainty-de-pendent trust—within a unified learning framework. Grounded in psychology and behavioral economics, it captures how individuals make probabilistic choices influenced by material payoffs, fear, trust, and social approval. Simulations of the Susceptible–Infected–Recovered process show that collective outcomes are best when agents display an intermediate level of rationality: they deliberate enough to respond to risk but remain flexible enough to adapt, avoiding the instability of both random and overly rigid decision-making. Regret exerts a dual influence—moderate levels encourage adaptive self-correction, while excessive regret or greed destabilize choices. Uncertainty has a similarly non-linear effect: moderate ambiguity promotes caution, but too much uncertainty disrupts coordination. Social norms restore cooperation by compensating for incomplete information. Personal norms guide behavior when individuals have reliable information and feel confident in their judgments. Injunctive norms—signals of others’ approval—become more influential under uncertainty, while descriptive norms, which arise from observing others’ actions, provide informational cues that help people decide what to do when direct knowledge is limited. Overall, the model provides a psychologically grounded, computationally explicit account of how emotion, cognition, and social norms jointly govern preventive behavior during epidemics.

\section{Introduction}
The persistence of vaccine hesitancy during recent epidemics shows that preventive behavior cannot be explained by instrumental rationality alone. Vaccination decisions arise from an interplay of material incentives, emotions, social expectations, and uncertainty about both disease dynamics and social reactions. Traditional epidemic and imitation-based models capture strategic and diffusion aspects but often overlook the psychological and normative forces driving real-world responses \cite{2016Bicchieri,2019Yamin}. Understanding how these forces interact is crucial for anticipating compliance and designing effective interventions.

Social norms—people’s perceptions of what others do (descriptive norms) and what others approve of (injunctive norms)—strongly shape vaccination behavior \cite{1990Cialdini,2016Bicchieri,2015Gelfand}. Descriptive norms guide behavior through informational cues and are most influential under uncertainty about others’ actions, whereas injunctive norms operate via moral and reputational channels \cite{2016Tankard}. Aligning these two norms enhances compliance, while misalignment undermines it \cite{2019Yamin,2008Gerber}. Large-scale studies show that providing accurate normative information—such as evidence of high vaccination rates—increases willingness to vaccinate, particularly among uncertain or mistrustful individuals \cite{2023Moehring}. What is more, evidence from experimental “pandemic-like” environments showed that individuals respond very differently to descriptive versus injunctive cues; in particular, Woike et al. \cite{2022Woike} found that only injunctive messages consistently reduced risk-taking, while descriptive messages could potentially even lead to backfiring, i.e. increasing risk-taking behavior. These findings underscore the importance of treating social norms as multidimensional influences rather than uniform informational signals, highlighting the need for models that capture how descriptive and injunctive expectations coevolve.

Emotions are equally central. Regret theory \cite{1982Loomes} shows that individuals evaluate outcomes relative to forgone alternatives, incorporating counterfactual emotions into choice. Anticipated regret increases vaccination uptake by raising the emotional cost of inaction \cite{2006Chapman,2016Brewer}, while worry mediates the link between perceived risk and compliance \cite{2020Hubner,2022Wong}. Thus, emotional forecasting and normative expectations jointly shape preventive behavior, motivating an integrated modeling approach.

Uncertainty compounds these processes. Individuals rarely know others’ infection or vaccination status, nor the reliability of their information sources. This imperfect information translates into subjective uncertainty—a lack of confidence in epidemiological and social beliefs. Empirical evidence shows that uncertainty—whether arising from incomplete social visibility or from low confidence in risk judgments—amplifies imitation and descriptive conformity while weakening injunctive guidance \cite{2025Zhuang}. Emotional arousal and unstable trust further modulate these effects \cite{2015Gelfand,2001Loewenstein}. Under uncertainty, decision weights shift dynamically across material, emotional, and normative domains. Our model formalizes this process through uncertainty- and fear-dependent weighting functions that bridge informational and psychological traditions \cite{2022Phelan,2020Hubner}.

Agent-based modeling (ABM) offers the ideal framework to integrate these mechanisms. ABMs represent heterogeneous, boundedly rational agents with emotional and social biases \cite{2022Czaplicka,2024Charalambous} interacting on structured networks \cite{2023Charalambous}. They reveal how local mechanisms—regret, conformity, trust—aggregate into collective equilibria and how feedback among emotion, information, and network structure drives norm evolution \cite{2013Conte,2022Andrighetto}. In vaccination contexts, ABMs have examined how social norms and perceived risks coevolve \cite{2024Charalambous} and how learning rules affect uptake \cite{2021Wu}. Combining behavioral experiments with simulation helps identify when societies transition between cooperation and defection \cite{2022Andrighetto}.

Here we integrate regret, uncertainty, and three-tier norm dynamics—personal, descriptive, and injunctive—within a unified agent-based behavioral epidemic model. The framework links bounded rationality, emotional feedback, and social expectations through a logit-based decision rule consistent with probabilistic choice under uncertainty \cite{1974McFadden,1999Camerer}. Individuals make probabilistic choices whose sensitivity to payoffs varies with fear, trust, and normative pressure \cite{2011Kahneman,2016Brewer,2023Moehring}, capturing how emotions bias preventive behavior and how low trust amplifies reliance on peer cues.

Existing behavioral-epidemic models typically treat regret, uncertainty, and norms in isolation, limiting their explanatory power. This work unifies them in a single, psychologically grounded computational framework. It contributes (i) a dynamic weighting scheme linking emotional and informational uncertainty to social influence, (ii) an additive regret term capturing affective self-correction, and (iii) coupled evolution equations for three interdependent norm types. Together, these elements contribute in clarifying how trust, emotion, and cognition coevolve under epidemic risk.

We address three central questions: (1) How do regret and uncertainty jointly shape vaccination decisions and epidemic outcomes? (2) Under what conditions do evolving social norms improve cooperation? (3) How does bounded rationality mediate the balance between self-correction and instability? By answering these, the model identifies when emotional and normative mechanisms promote resilience—or drive breakdown—in collective preventive behavior.

\section{Model}

During a pandemic, two coupled processes unfold in parallel: the epidemiological spread of infection and the behavioral adaptation of individuals responding to it. Because each shapes the other, they must be modeled jointly. Disease transmission occurs on a physical-contact network following the classical Susceptible--Infected--Recovered (SIR) framework, while behavioral adaptation evolves on a social network where individuals observe and influence one another.

\begin{figure}
\centering
\begin{minipage}[t]{0.45\columnwidth}%
\vspace{0pt}
\begin{algorithm}[H]
\caption{Evolutionary Update Rule}
\begin{algorithmic}[1]
\State \textbf{Input:} Networks $G^{Phys}$,$G^{Soc}$, SIR $(\beta,\mu)$, Utility $(c_V,c_I,\kappa,m,\eta_1,\eta_2)$, Norms $(y_i,\tilde{x}_i,\tilde{y}_i)$
\State \textbf{Output:} $I(t)$, $x_i(t)$, $(y_i,\tilde{x}_i,\tilde{y}_i)$
\State Initialize norms and SIR state.
\For{season $t=0,1,2,\ldots$}
  \State Run $n_{\mathrm{sim}}$ SIR realizations; compute $E_i[P](t)$, $S_i(t)$, $f_i(t)$, $\hat I_i(t)$.
  \State Learning: update $\hat\pi^{\text{Unvac}}_i(t)$ via Eq.~(21); set $\hat\pi^{\text{Vac}}_i=1-c_V$.
  \State Regret (LS): $R(x)=\eta_1 x^{\eta_2}$ if $x>0$ else $0$; adjust $\pi^{\text{Vac}}_i,\pi^{\text{Unvac}}_i$.
  \State Weights: compute $u_i,q_i^{\text{stab}}$, $q_i^{\text{cons}}$; then $A_{\text{mat}},A_y,A_{\tilde{x}},A_{\tilde{y}}$.
  \State Utilities: $\Delta U_i$ from Eq.~(4); intention $x_i=(1+e^{-\Delta U_i/\kappa})^{-1}$; draw $a_i$.
  \State Norm dynamics: update $(y_i,\tilde{y}_i,\tilde{x}_i)$ via Eqs.~(23)--(26).
  \State If $\lvert\langle x_i\rangle_{t+1}-\langle x_i\rangle_t\rvert<\varepsilon$ for 50 seasons: \textbf{break}.
\EndFor
\end{algorithmic}
\end{algorithm}
\end{minipage}%
\hfill
\begin{minipage}[t]{0.5\columnwidth}%
\vspace{2.0em}
\centering
\includegraphics[width=1.05\linewidth]{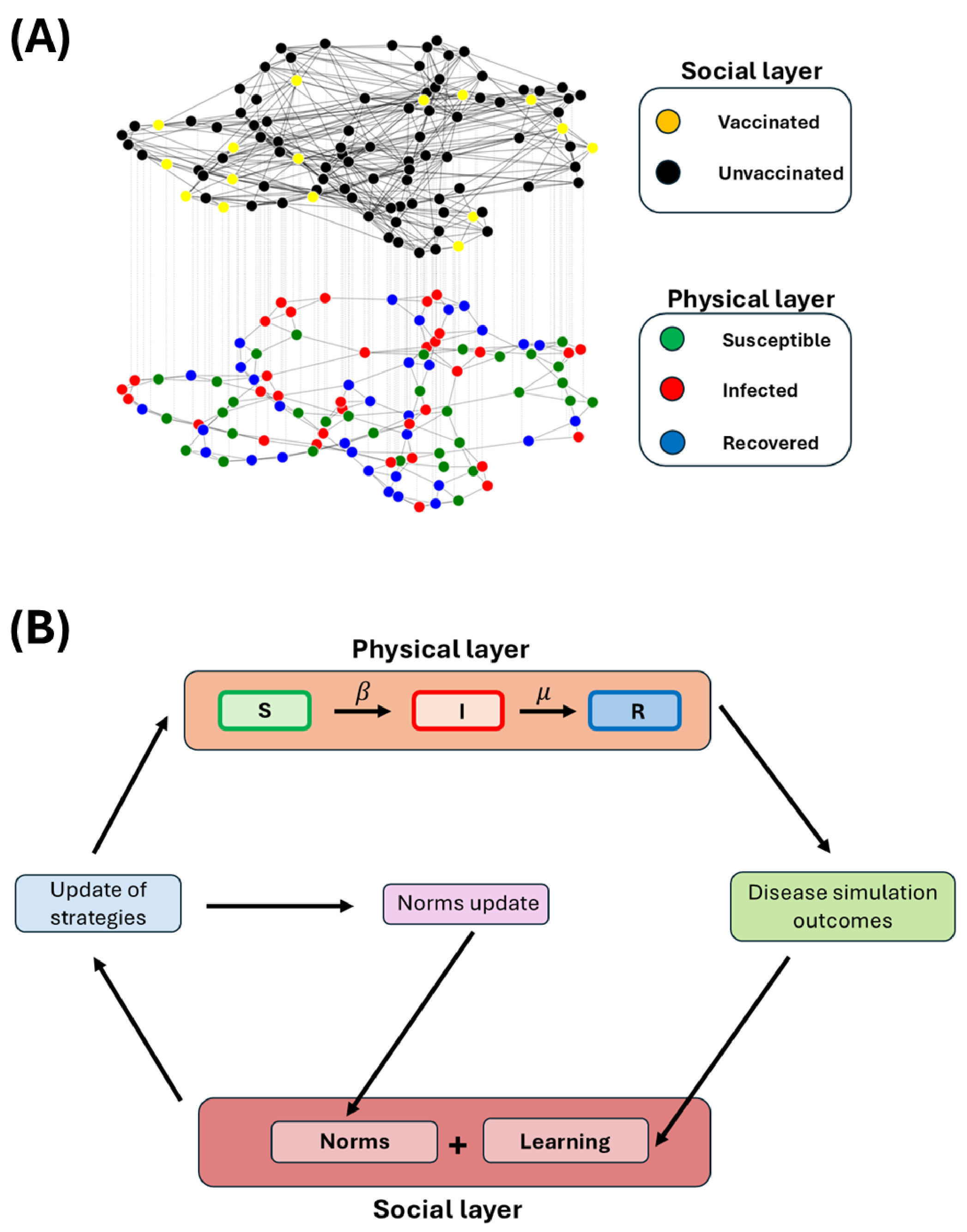}

\vspace{3.5em}

\end{minipage}

\caption{\label{fig:Fig1}
\textbf{Left: Algorithm of system's dynamics.}  
The flowchart outlines the sequence of processes within each epidemic season. The model first simulates infection dynamics on the physical layer, estimating each agent’s infection probability. After each season, agents decide whether to vaccinate based on the previous outcomes, integrating both material (learning-based) and normative considerations. Subsequently, they update their norms before the next season begins. \textbf{Right (A): Schematic of the two layer multiplex network.} The model operates on two interconnected layers. On the \textit{physical layer}, infection propagates through a Susceptible–Infected–Recovered (SIR) process; on the \textit{social layer}, agents observe outcomes and peers’ actions to guide vaccination. The physical layer follows a small-world topology, while the social layer uses a Klimek–Thurner structure—both empirically supported as realistic for contact and communication networks. A significant overlap between layers reflects that many physical ties also transmit social information. 
\textbf{(B): Dynamics of the vaccination game.}  Schematic of the algorithm. Each season, the SIR dynamics run on the physical layer to estimate infection probabilities. Agents then decide whether to vaccinate, considering both payoff-based learning and normative factors, update their norms accordingly, and begin the next season. 
}
\end{figure}

Although network topology is not the primary focus, we employ empirically grounded synthetic structures (Fig.~\ref{fig:Fig1}). The physical layer $G^{Phys}$ follows a small-world network \cite{2010Salathe} with average degree 6 and rewiring probability 0.1, and the social layer $G^{Soc}$ is generated using the Klimek--Thurner model \cite{2013Klimek} (with parameters $r=0.12$, $c=0.58$ and $m=1$). To reflect that many physical interactions also convey information, we assume partial overlap between the two layers. Tests with alternative topologies, such as Erdős--Rényi graphs, yield qualitatively consistent results.

At the start of each season, agents decide whether to vaccinate, and one individual is initialized as infected. The process repeats across seasons until equilibrium is reached. Within each season, multiple SIR realizations are simulated using an event-driven algorithm \cite{2017Kiss} to estimate infection probabilities. The transmission and recovery rates are denoted by $\beta$ and $\mu$; since $\mu$ only determines timescale, it is normalized to $\mu=1$.

Behavioral adaptation is modeled game-theoretically, as each individual’s payoff depends on others’ choices. Vaccination decisions are influenced not only by perceived disease risk but also by beliefs and social norms. Following \cite{2021Gavrilets}, each individual $i$ possesses three normative components: (i) a personal attitude $y_i$ reflecting moral preference; (ii) an empirical expectation $\widetilde{x}_i$ representing perceived peer behavior; and (iii) a normative expectation $\widetilde{y}_i$ capturing perceived collective approval.

Focusing on seasonal epidemics, we integrate information from repeated SIR simulations with agents’ evolving norms to determine vaccination choices at each new season. Rather than continuously tracking norm change, we update all normative variables only at decision points—the start of each season—capturing how experience and feedback shape subsequent behavior. The complete algorithmic structure of this coupled epidemiological--behavioral system is illustrated in Fig.~\ref{fig:Fig1}.

\subsection{Decision making process}
To evaluate vaccination intentions, we define a utility function integrating material, psychological, and normative components. The \textit{Regret--Uncertainty model} combines three empirically grounded mechanisms---anticipated regret, social norms, and uncertainty-dependent trust---into a unified decision framework. Preventive behavior thus reflects not only instrumental payoffs but also affective and social motives, especially under uncertainty and interdependence.

An agent’s intention $x_i$ depends on the difference between the utilities of the two available actions. The probability of vaccination increases with this difference:
\begin{equation}
U_i(\pi_{Vac}, \pi_{Unvac}, a, y, \tilde{x}, \tilde{y})
= A_{mat}\,\Pi_i(\pi_{Vac}, \pi_{Unvac}, a)
+ V_i(a, y, \tilde{x}, \tilde{y}),
\end{equation}
where $a \in \{0,1\}$ represents the action (0 = not vaccinate, 1 = vaccinate), $y$ is the personal norm, and $\tilde{x}, \tilde{y}$ are the empirical and normative expectations respectively. The first term captures material considerations weighted by $A_{mat}$; the second term reflects psychological and social influences.

\textbf{Material utility:}
\begin{equation}
\Pi_i(\pi_{Vac}, \pi_{Unvac}, a) = a\pi_{Vac} + (1 - a)\pi_{Unvac},
\end{equation}
where $\pi_{Vac}$ and $\pi_{Unvac}$ are average payoffs from vaccinating or not over the last $m$ seasons.

\textbf{Normative utility} \cite{2022Tverskoi}:
\begin{equation}
V_i(a, y, \tilde{x}, \tilde{y}) =
A_y |a - y| + k_a |1 - a - \tilde{x}| - k_d |a - \tilde{x}|
+ A_{\tilde{y}}(2\tilde{y} - 1)(a - 0.5),
\end{equation}
where $A_y, k_a, k_d,$ and $A_{\tilde{y}}$ quantify the strengths of:
\begin{enumerate}
    \item \textbf{Cognitive dissonance}: mismatch between behavior and moral belief $A_y|a-y|$;  
    \item \textbf{Descriptive conformity}: alignment with peers’ actions $k_a|1-a-\tilde{x}| - k_d|a-\tilde{x}|$; Parameters $k_a$ and $k_d$ characterize the psychic benefit of approval and the costs of disapproval by peers, respectively, as in \cite{2022Tverskoi}. 
    \item \textbf{Injunctive conformity}: sensitivity to perceived moral approval $A_{\tilde{y}}(2\tilde{y}\!-\!1)(a\!-\!0.5)$
\end{enumerate}

This formulation integrates cognitive dissonance and social conformity into a single additive structure, reflecting how moral coherence and peer alignment jointly shape utility.

Following \cite{2022Tverskoi}, the difference in utilities between vaccinating and not vaccinating is:
\begin{equation}
\Delta U_i = U_i(1) - U_i(0)
= A \!\left[\frac{2}{A}\!\left(A_{mat}p_{learn} + A_y y + A_{\tilde{x}}\tilde{x} + A_{\tilde{y}}\tilde{y}\right) - 1\right],
\end{equation}
where
\begin{equation}
p_{learn} = 0.5 + 0.5(\pi_{Vac} - \pi_{Unvac}), \qquad
A = A_{mat} + A_y + A_{\tilde{x}} + A_{\tilde{y}}.
\end{equation}

\paragraph{Bounded Rationality}

The probability that agent $i$ vaccinates follows a Fermi-type function:
\begin{equation}
x_i = \frac{1}{1 + e^{-\Delta U_i / k_{rat}}},
\end{equation}
where $k_{rat}$ is the rationality parameter. This \textit{logit quantal-response} formulation implies probabilistic decision-making: as $\Delta U_i$ grows, vaccination becomes more likely. Without loss of generality, we set $A=1$. The coefficients $A_{mat}$, $A_y$, $A_{\tilde{x}}$, and $A_{\tilde{y}}$ denote the adaptive weights of material, personal, descriptive, and injunctive components, respectively, whose evolution is described below.

\subsubsection{Dynamical Weights}

To determine how each component contributes to decision-making, we rely on empirical findings linking normative influence to psychological states and environmental stability.

Studies show that as infection risk increases, individuals rely more on descriptive than injunctive norms \cite{2023Heiman}, and uncertainty about susceptibility amplifies the motivational power of descriptive norms \cite{2025Zhuang}. Under fear, people seek more social information \cite{2023WuOuyang}, and perceived knowledge insufficiency fosters imitation and peer reliance \cite{2020Hubner}. However, such dependence presupposes that peers are stable and relatively consistent in their behavior.

Integrating these findings, we define:
\begin{align}
A_{mat}(\phi_i^{Emp}, \phi_i^{Col}) &= (1 - \phi_i^{Emp})(1 - \phi_i^{Col}),\\
A_y(\phi_i^{Emp}, \theta_i^{Col}) &= \phi_i^{Emp}(1 - \theta_i^{Col}),\\
k_a + k_d = A_{\tilde{x}}(\phi_i^{Emp}, \phi_i^{Col}) &= (1 - \phi_i^{Emp})\phi_i^{Col},\\
A_{\tilde{y}}(\phi_i^{Emp}, \theta_i^{Col}) &= \phi_i^{Emp}\theta_i^{Col},
\end{align}
which satisfy $A = 1$ and $A_j \in [0,1]$ for all $j$. In addition, $k_a+k_d$ can be viewed as a measure of the strength of injunctive social norms. Here, $\phi_i^{Emp}$ balances empirical and injunctive influence, while $\phi_i^{Col}$ (the \textit{trust coefficient}) governs the weight assigned to collective versus individual cues:
\begin{equation}
\phi_i^{Emp} = (f_i u_i)^{1/2}, \quad
\phi_i^{Col} = (q_i^{stab} q_i^{cons} f_i u_i)^{1/4}, \quad
\theta_i^{Col} = (q_i^{stab} q_i^{cons} f_i u_i^{intr})^{1/4}.
\end{equation}

Here $f_i = 1 - S_i(t)$ quantifies fear, increasing as perceived safety $S_i(t)$ decreases:
\begin{equation}
S_i(t) =
\begin{cases}
1 - \widehat{I}_i(t), & \text{if unvaccinated},\\
1 - E_i[P](t), & \text{if vaccinated},
\end{cases}
\end{equation}
where $\widehat{I}_i(t) = n_{sim}^{inf}(t)/n_{sim}$ and $E_i[P](t)$ is the expected infection probability derived from infected neighbors.

Total uncertainty is
\begin{equation}
u_i = u_i^{intr} + u_i^{info},
\end{equation}
where $u_i^{intr}$ represents intrinsic uncertainty (lack of confidence) and $u_i^{info}$ captures informational uncertainty arising from incomplete knowledge of neighbors’ infection states.

The variable $q_i^{stab}$ quantifies peer stability based on the change-detector function \cite{2007Ho}:
\[
q_i^{stab}(t) = 1 - \frac{1}{2} Q_i(t), \quad Q_i(t) = \sum_{k=1}^{m-1}(h_i^k(t) - q_i^k(t)),
\]
where $h_i^k(t)$ measures historical frequencies of neighbor actions over a memory window $m$, and $q_i^k(t)$ records current strategies. High volatility (large $Q_i$) decreases $q_i^{stab}$, prompting agents to rely more on self-learning.

Local consensus is given by $q_i^{cons} = 2|X_i - 0.5|$, where $X_i$ is the fraction of vaccinated neighbors. Thus, $q_i^{cons}=0$ when opinions are evenly split and $q_i^{cons}=1$ when full agreement exists.

The equations define the weights as geometric means of fear, trust, and uncertainty—an appropriate formulation when independent factors jointly determine the reliability of social information. This multiplicative rule ensures that if any component is weak, overall influence declines sharply, consistent with evidence that (i) fear without trust leads to disengagement \cite{2024Yewell}, (ii) trust without fear lacks motivational urgency \cite{2020Hubner}, and (iii) uncertainty without credible input results in inaction.  

\begin{figure}[H]
\begin{centering}
\begin{tabular}{cc}
\includegraphics[scale=0.265]{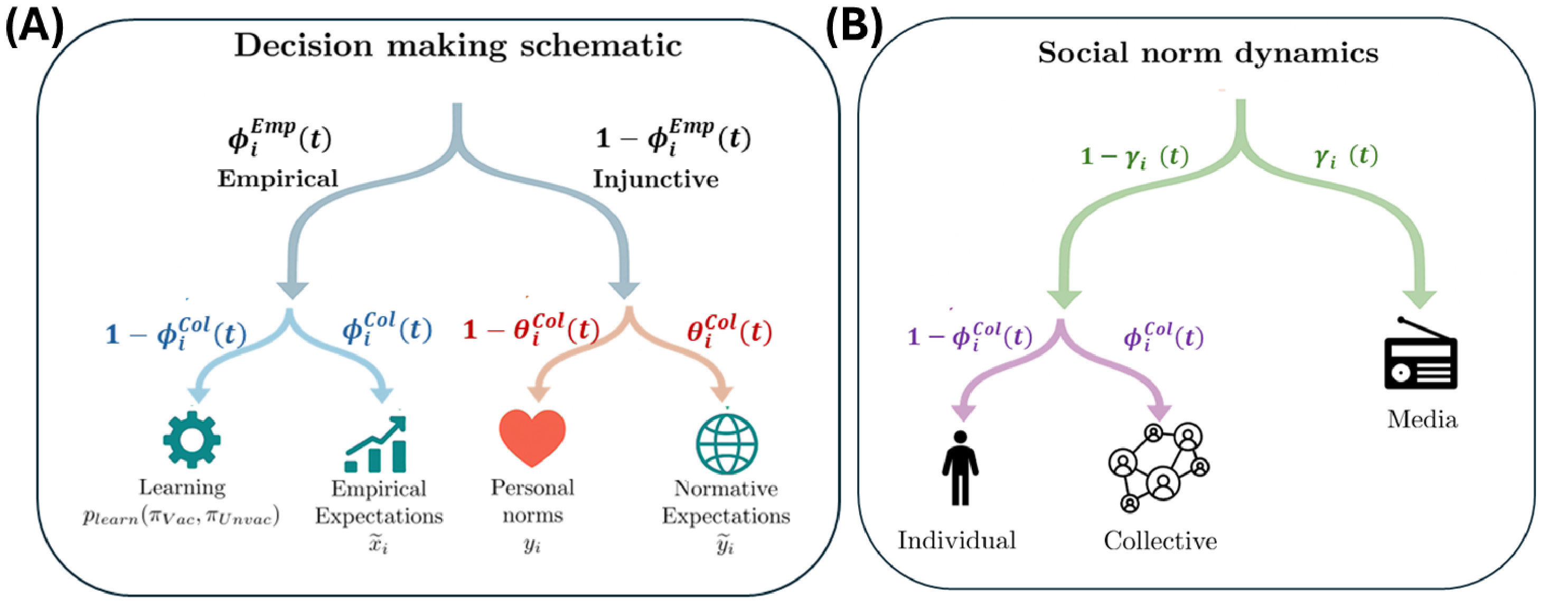}
\tabularnewline
\end{tabular}
\par\end{centering}
\caption{\label{fig:Fig2} \textbf{(A): Schematic of the decision-making process.}  
The diagram illustrates how learning and social norms jointly determine each agent’s vaccination intention, weighted by their relative influence. Two assumptions guide this interaction:  
(a) The less safe an agent feels—i.e., the lower $S_{i}(t)$, representing infection risk from the previous season—the more they rely on empirical cues (learning or empirical expectations) rather than normative ones (personal beliefs or injunctive expectations). Similarly, the more uncertain the agent feels about the knowledge he posses, the more he will consult his injunctive, moral variables (personal and normative expectations). 
(b) The more heterogeneous and unstable the environment—when agents’ recent decisions diverge from one another or from past averages—the greater the reliance on personal experience and beliefs instead of social imitation. \textbf{(B): Schematic of the dynamics of social norm variables.}  
The diagram shows how different factors shape norm evolution under two key assumptions:  
(a) Each agent updates her norms with probability $\gamma_{i}$ under external influence, otherwise combining personal beliefs, existing norms, and peers’ actions.  
(b) Environmental instability—large deviations between recent and past behaviors and low consensus among neighbors—increases dependence on personal factors over conformity to others.
}
\end{figure}

By combining these drivers multiplicatively, the model captures how individuals integrate emotional, cognitive, and social cues under risk. Fear heightens perceived vulnerability, trust legitimizes information, and uncertainty fosters openness to influence. Together, these mechanisms determine how descriptive and injunctive norms guide vaccination behavior. The overall decision process is summarized schematically in Fig.~\ref{fig:Fig2}.

\subsubsection{Uncertainty}


During COVID-19, countries adopted markedly different digital contact-tracing systems. Centralized platforms such as China’s \textit{Alipay Health Code} and South Korea’s \textit{Corona 100 m} relied on government databases to provide real-time information about contacts with infected, non-infected, or untested individuals \cite{2022Yu,2020Liang,2021Zhou,2021Cong,2020Ryan}. In contrast, decentralized approaches like the MIT-led \textit{SafePaths} project stored contact histories locally and compared them with anonymized case data, alerting users to potential exposure without revealing identities \cite{2021Baker,2024Muntoni,2020Raskar,2020Tarkoma,2021Elkhodr,2021Bianconi}. Although such systems typically omitted encounters with non-infected or untested persons, expanding their functionality to include these would be technically feasible.

Building on this context, we model epidemic dynamics under partial observability of the social network. Rather than assuming that individuals know the infection status of all contacts, we posit that each agent observes only a subset of neighbors—reflecting the limited information available through digital tracing tools. This incomplete visibility introduces uncertainty about true infection risk.

To quantify uncertainty, we compute each agent’s expected infection probability $E_i[P]$ and its variance $\mathrm{Var}_i[P]$, which jointly reflect perceived risk and confidence. Let $n_i^{\text{total}}$ denote the total number of neighbors of agent $i$, and $n_i^{\text{obs}} = z\, n_i^{\text{total}}$ the subset whose infection state is known, where $z$ represents the observed fraction. Among the observed neighbors,
\begin{equation}
k_i^{\text{obs}} = n_i^{\text{obs}}\, \widehat{I}_{i,\text{neighbors}}(t),
\end{equation}
with $\widehat{I}_{i,\text{neighbors}}(t) = n_{\text{sim}}^{\text{inf,neigh}}(t)/n_{\text{sim}}$ denoting the average proportion of infected neighbors across simulation runs.  

The remaining $n_i^{\text{unobs}} = n_i^{\text{total}} - n_i^{\text{obs}}$ neighbors have unknown states and are assumed independently infected with probability $p$. The number of infected unobserved neighbors then follows
\[
X \sim \mathrm{Binomial}(n_i^{\text{unobs}}, p).
\]
For a realization $x$ of $X$, the total number of infected contacts is $k = k_i^{\text{obs}} + x$. Assuming a per-contact transmission probability $\beta = 1 - e^{-\lambda}$, the infection probability conditional on $k$ infected contacts is
\[
P_k = 1 - (1 - \beta)^k.
\]

The expected infection probability is
\[
E_i[P] = \sum_{x=0}^{n_i^{\text{unobs}}} \Pr(X=x)\,\bigl[1-(1-\beta)^{k_i^{\text{obs}}+x}\bigr],
\]
where $\Pr(X=x)$ is the corresponding binomial probability.  
The second moment and variance are
\[
E_i[P^2] = \sum_{x=0}^{n_i^{\text{unobs}}} \Pr(X=x)\,\bigl[1-(1-\beta)^{k_i^{\text{obs}}+x}\bigr]^2, 
\qquad
\mathrm{Var}_i[P] = E_i[P^2] - (E_i[P])^2.
\]

We define the informational uncertainty of susceptibility as
\[
u_i^{\text{info}} = \frac{\mathrm{Var}_i[P]}{\mathrm{Var}_{\max}} ,
\]
where $\mathrm{Var}_{\max}$ is the variance when $k_i^{\text{obs}} = 0$ (no known neighbor states), ensuring $0 < u_i^{\text{info}} < 1$.

This formulation captures the ambiguity in risk assessment created by incomplete information—especially relevant in privacy-preserving, decentralized tracing systems where neighborhood states are only partially visible. By explicitly accounting for unobserved contacts, the model links epidemiological dynamics to information constraints in digital health technologies.

\subsection{Experience dependent mechanism}


Human behaviour reflects a mix of model-free and model-based reinforcement learning \cite{1999Camerer}, combining accumulated experience with forward-looking considerations. The Experience-Weighted Attraction (EWA) framework captures this integration \cite{2013Galla}, bringing together reinforcement learning \cite{2018Sutton} and belief learning \cite{2012Feltovich} by assigning equal weight to realized and forgone payoffs. In our setting, forgone payoffs are inferred from neighbours’ average infection rates because individuals do not observe others’ outcomes directly \cite{2018Sanchez}. Agents evaluate payoffs over the past $m$ cycles, with memory decaying at a rate determined by perceived epidemiological risk. EWA thus describes how individuals merge past experience with beliefs about others’ behaviour. Its empirical support \cite{2018RealpeGomez} makes it a suitable, cognitively grounded foundation for modelling social adaptation and norm compliance.

At the end of each season, agents evaluate the payoffs associated with the two available actions—vaccinate or not vaccinate:  
\[
\begin{array}{c}
\Pi_i^{Unvac}(t) =
\begin{cases}
1 - c_I E_i[P](t), & \text{if vaccinated},\\
1 - c_I \widehat{I}_i(t), & \text{if unvaccinated},
\end{cases}\\
\Pi_i^{Vac}(t) = 1 - c_V,
\end{array}
\]
where $c_V$ and $c_I$ denote the costs of vaccination and infection, respectively.

\paragraph{Memory Decay}

Let $\widehat{\pi}_i^{Vac}(t)$ and $\widehat{\pi}_i^{Unvac}(t)$ represent the average payoffs of the two actions over the past $m$ seasons. Since the payoff from vaccination is constant,  
\[
\widehat{\pi}_i^{Vac}(t) = \Pi_i^{Vac} = 1 - c_V.
\]
For the unvaccinated option, memory decays over time depending on the agent’s emotional state. When perceived safety $S_i(t)$ is low, fear increases attention to recent outcomes, causing older experiences to be discounted more strongly:  
\[
\widehat{\pi}_i^{Unvac}(t) =
\sum_{j=0}^{m}
\frac{(S_i(t))^{j}}{\sum_{n=0}^{m} (S_i(t))^{n}}
\Pi_i^{Unvac}(t-j),
\]
where $S_i(t)$ is the safety parameter defined previously.

\subsubsection*{Regret}

Anticipated regret is a key driver of preventive behavior. Empirical studies on influenza and COVID-19 vaccination show that regret and worry mediate the link between perceived risk and uptake, shaping both immediate and future choices \cite{2006Chapman,1995Richard,2004Wroe,2022Wong}. Regret thus operates as an experience-weighted feedback mechanism—past outcomes influence future intentions \cite{2012Wells}.

To formalize this, we employ the Loomes--Sugden (LS) model of regret \cite{1982Loomes,1982Bell}, which introduces an additive regret--rejoice term to utility without conflating it with risk aversion. The payoff adjusted for regret is  
\[
\pi_{Vac} = \widehat{\pi}_{Vac} - R(\widehat{\pi}_{Unvac} - \widehat{\pi}_{Vac}),
\]
and symmetrically for $\pi_{Unvac}$, where
\[
R(x) =
\begin{cases}
\eta_1 x^{\eta_2}, & x > 0,\\
0, & x \leq 0,
\end{cases}
\]
with $\eta_1$ and $\eta_2$ denoting the strength and curvature of regret. When an inferior option is chosen, the resulting utility is reduced by the regret of not selecting the better alternative.

The LS model provides a parsimonious, empirically supported account of emotional feedback. It captures how anticipated regret promotes immediate vaccination, whereas experienced regret shapes future intentions \cite{2006Chapman,2007Zeelenberg}. Unlike Prospect Theory, which emphasizes perceptual biases such as loss aversion and probability weighting \cite{2024Li}, the LS formulation isolates regret’s motivational role as a corrective, counterfactual mechanism—well suited to repeated, feedback-driven settings like seasonal vaccination \cite{2015Bleichrodt}.

We employ the LS model because it reproduces a key empirical regularity: anticipated regret predicts vaccination more reliably than perceived risk, while experienced regret updates subsequent behavior. It does so with minimal parameters and clear interpretability, directly linking emotional self-evaluation to learning and adaptation.

Although \textit{normative regret} (e.g., guilt or social disapproval) could be modeled similarly, evidence indicates these emotions operate mainly ex post, reinforcing self-correction and social alignment rather than foresight \cite{2007Zeelenberg,2007Baumeister}. Accordingly, we restrict regret here to material outcomes and introduce normative regret, i.e. social disapproval, in the next section on norm dynamics.

\subsection{Norm dynamics}


To integrate social norms into the Experience-Weighted Attraction (EWA) framework, we separate material payoffs from normative dynamics. Empirical research identifies three complementary influences—descriptive norms (what others do), injunctive norms (what others approve of), and personal norms (one’s moral standard) \cite{1990Cialdini,2005Bicchieri}. These operate through distinct cognitive pathways and evolve on different timescales. Longitudinal evidence from the Swiss COVID-19 vaccination campaign shows that media tone primarily shaped injunctive norms, whereas descriptive norms closely tracked observed vaccination behavior \cite{2024Geber}. Furthermore, online experiments suggest that when social information is limited or unreliable, personal norms best predict behavior \cite{2021Catola}. With such empirical evidence in mind, our model tracks the joint evolution of $y_i$, $\tilde{x}_i$, and $\tilde{y}_i$, capturing both social learning and moral feedback.

Agents are boundedly rational \cite{1957Simon}: they infer others’ attitudes from observed actions \cite{2002Hedden,2021Szekely}, balancing the avoidance of social disapproval with the need for internal coherence. The three normative variables are updated after each decision cycle, distinguishing between the discrete vaccination action and the continuous intention $x_i$—unlike \cite{2021Gavrilets}, where actions were modeled as continuous.

The dynamics of personal norms ($y_i$), injunctive expectations ($\tilde{y}_i$), and descriptive expectations ($\tilde{x}_i$) follow a DeGroot-type formulation \cite{2021Gavrilets}. The corresponding terms represent cognitive dissonance (reducing action–attitude mismatch), social projection (assuming similarity to others \cite{2007Krueger}), logical consistency between beliefs and observed actions \cite{2016Friedkin}, and peer or authority conformity \cite{2015Kashima}. Dissonance aligns $x_i$ and $y_i$, ensuring that attitudes adjust toward past behavior \cite{1994Rabin}.

For interpretability, we normalize these equations:
\begin{equation}
\begin{array}{c}
y'_i = y_i + \xi_i^1[\widehat{C_i^{11}}x_i + \widehat{C_i^{12}}X_i + \widehat{C_i^{13}}G_i^1 - y_i]\\
\tilde{y}'_i = \tilde{y}_i + \xi_i^2[\widehat{C_i^{21}}y_i + \widehat{C_i^{22}}X_i + \widehat{C_i^{23}}G_i^2 - \tilde{y}_i]\\
\tilde{x}'_i = \tilde{x}_i + \xi_i^3[\widehat{C_i^{31}}\tilde{y}_i + \widehat{C_i^{32}}X_i + \widehat{C_i^{33}}G_i^3 - \tilde{x}_i],
\end{array}
\label{eq:NormDynamics}
\end{equation}
where $\xi_i^q = \sum_j C_i^{qj}$ denotes each variable’s updating rate ($q=1,2,3$), and $\widehat{C_i^{qj}} = C_i^{qj}/\sum_k C_i^{qk}$ ensures $\sum_j \widehat{C_i^{qj}}=1$.

\paragraph{Influence Weights and Trust}

We initially neglect external authorities by setting $\widehat{C_i^{q3}}=0$, implying $\widehat{C_i^{q2}}=1-\widehat{C_i^{q1}}$. When external interventions are included, they are uniform across agents: $\widehat{C_i^{q3}}=\gamma_q$. Each agent’s update thus depends on internal consistency ($x_i$, $y_i$, $\tilde{y}_i$) and social influence ($X_i$). We interpret $\widehat{C_i^{q1}}$ as the self-consistency weight and $1-\widehat{C_i^{q1}}$ as the social influence weight. These coefficients are coupled to the epidemic environment through the trust factor $\theta_i^{Col}(t)$, derived earlier:
\[
\widehat{C_i^{q1}} = 1 - \theta_i^{Col}(t), \qquad
\widehat{C_i^{q2}} = \theta_i^{Col}(t),
\]
where $\theta_i^{Col}(t)$ depends on the change-detector and consensus functions defined previously. Figure~\ref{fig:Fig2} schematically illustrates these relationships.

Unless stated otherwise, we set $\xi_i^1 = 0.01$, $\xi_i^2 = 0.1$, and $\xi_i^3 = 1$, reflecting that personal norms evolve more slowly than injunctive or descriptive expectations ($\xi^3>\xi^2>\xi^1$). Results remain robust as long as this hierarchy is preserved. This hierarchy is experimentally justified as evidence shows that personal norms remain stable and predictive even when empirical and normative expectations vary, suggesting that personal moral evaluations are less context-dependent than descriptive or injunctive cues \cite{2021Catola}. Future extensions could introduce state-dependent $\xi_i^q$, potentially yielding qualitative transitions similar to those reported in \cite{2024Vriens}.

%

\section{Results}

We ran 1,000 simulations of the Susceptible–Infected–Recovered (SIR) process on a network of $N=500$ nodes, using the event-driven algorithm of \cite{2017Kiss} each season. Vaccination and infection costs were set to $c_V = 1$ and $c_I = 0.1$, with intrinsic susceptibility uncertainty $u_i^{intr} = 0.1$ (results were robust for nearby values). Unless stated otherwise, memory length was $m=4$ and transmission rate $\beta = 6$. Initial personal, descriptive, and injunctive norms were drawn independently from uniform distributions. Each run started with a single infected agent and no initial vaccinations, continuing until equilibrium—defined as a vaccination-rate change below $0.01$ during the last 50 of a maximum 200 seasons. Reported outcomes are medians with interquartile ranges.

Robustness checks on Erdős–Rényi and scale-free networks with average degree $\langle k \rangle = 6$ showed variations in vaccination and infection levels below 5\%, indicating that results depend mainly on clustering and degree heterogeneity rather than specific topology.

Parameter choices were guided by empirical evidence from influenza and COVID-19 vaccination studies and experiments on regret and probabilistic choice. Anticipated regret is a strong emotional predictor of vaccination: a one–standard-deviation increase typically raises intention by 10–25\% \cite{2006Chapman,2016Brewer,2022Wong}. Setting the regret-sensitivity coefficient to $\eta_1 \approx 1$ on a normalized utility scale reproduces this elasticity, while $\eta_2 \in [0.5,1]$ captures moderate self-serving bias, consistent with experimental estimates of regret concavity \cite{2015Bleichrodt,2007Zeelenberg}.

The rationality parameter $\kappa_{\mathrm{rat}}$ governs stochasticity in logit choice. Values between 0.1 and 1 correspond to precision levels $\lambda = 1/\kappa_{\mathrm{rat}} \in [1,10]$, matching quantal-response estimates from coordination and strategic games \cite{1999Camerer,2001Goeree}. Within this range, agents select the higher-utility option with 65–90\% probability, consistent with observed variability in preventive behavior. Parameters were not fitted but chosen within empirically grounded bounds to capture realistic, probabilistic, and norm-sensitive dynamics.

\subsection{Social norms evolution}

\begin{figure}[th!]
\centering

\includegraphics[scale=0.325]{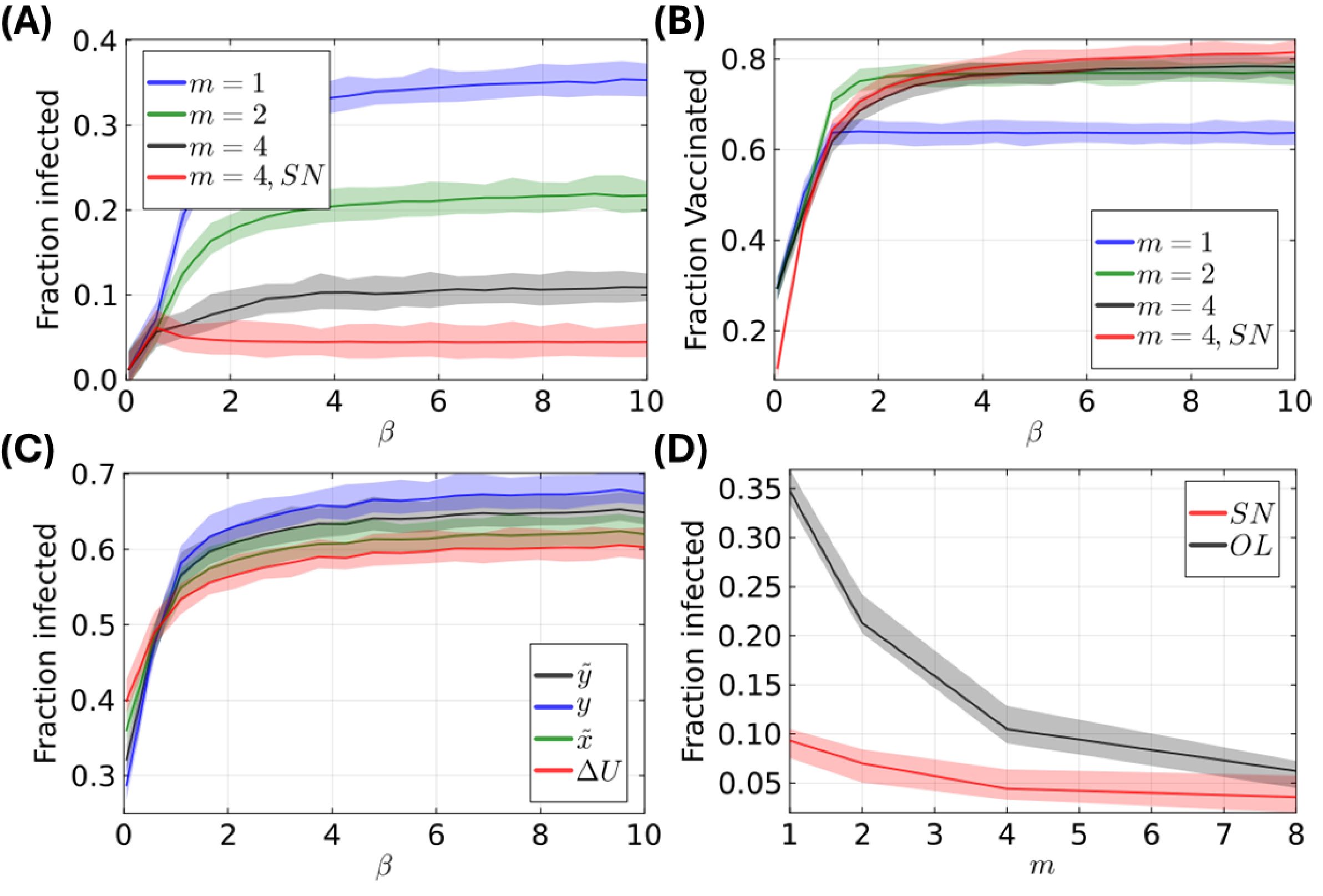}
%
%

\caption{\label{fig:Fig3}
\textbf{Behavioral and epidemiological effects of memory and social norms.}
\textbf{(A)} Fraction of infected individuals vs.\ infectivity rate $\beta$.  
Without social norms, increasing $\beta$ raises infection levels. Comparing memory lengths $m=1$ (blue), $m=2$ (green), and $m=4$ (black), longer memory—hence greater informational depth—consistently lowers infections. When agents also consider perceived norms (red), infections drop further for the same memory, indicating that normative feedback stabilizes preventive behavior.
\textbf{(B)} Fraction of vaccinated individuals vs.\ infectivity rate $\beta$.  
Vaccination coverage rises with both $\beta$ and memory but quickly saturates. Including social norms ($m=4$) does not increase mean vaccination relative to learning-only agents yet improves its spatial distribution. Norms therefore enhance coordination and reduce clustering rather than overall uptake ($k_{\mathrm{rat}}=0.1$).
\textbf{(C)} Evolution of social norms vs.\ infectivity rate $\beta$.  
The intention $\Delta U_i$ and the three norm components remain misaligned because they adapt at different rates: empirical expectations adjust fastest, while personal and injunctive norms evolve more slowly ($k_{\mathrm{rat}}=0.1$).
\textbf{(D)} Fraction of infected individuals vs.\ memory length.  
With only learning (OL), infections decline sharply as memory increases, reflecting improved decision accuracy. When social norms (SN) are included, infection rates become less sensitive to memory, showing that normative feedback substitutes for experience. Social influence thus reduces reliance on individual memory to sustain cooperative vaccination.}
\end{figure}

Figures~\ref{fig:Fig3}A and~\ref{fig:Fig3}B show the joint evolution of epidemic and behavioral dynamics under varying cognitive and normative conditions. In Fig.~\ref{fig:Fig3}, infection levels rise with infectivity $\beta$, but both longer memory and social norms markedly reduce outbreak size. Agents with extended memory ($m=4$ vs.\ 1–2) incorporate more payoff history, vaccinate more consistently, and face smaller epidemics. Adding social norms (red line) further suppresses infections for the same memory length, even when average vaccination coverage changes little. Norms therefore improve the \textit{distribution} rather than the \textit{level} of vaccination—reducing clustering and stabilizing uptake. 

Figure~\ref{fig:Fig3}C depicts the evolution of personal norms ($y$), empirical expectations ($\tilde{x}$), injunctive expectations ($\tilde{y}$), and behavioral intentions ($x$). These variables remain distinct because they evolve at different speeds: empirical expectations react most rapidly to observed behavior, injunctive norms adjust more slowly through social feedback, and personal norms change gradually through reflection. This temporal hierarchy prevents convergence and captures the asymmetry described by \cite{2005Bicchieri,2014Oraby}, where descriptive norms follow short-term cues while injunctive norms act as slower moral anchors stabilizing behavior under changing risk.

Finally, Fig.~\ref{fig:Fig3}D illustrates how memory length influences vaccination dynamics. Longer recall produces steadier vaccination and smaller outbreaks, as extended memory dampens short-term fluctuations. When social norms are included, the marginal effect of memory weakens—normative feedback already promotes behavioral stability. Beyond moderate memory ($m \ge 4$), gains plateau, indicating that limited recall suffices to sustain cooperative vaccination.

\subsubsection{Bounded rationality}

\begin{figure}[h!]

\begin{centering}
\includegraphics[scale=0.32]{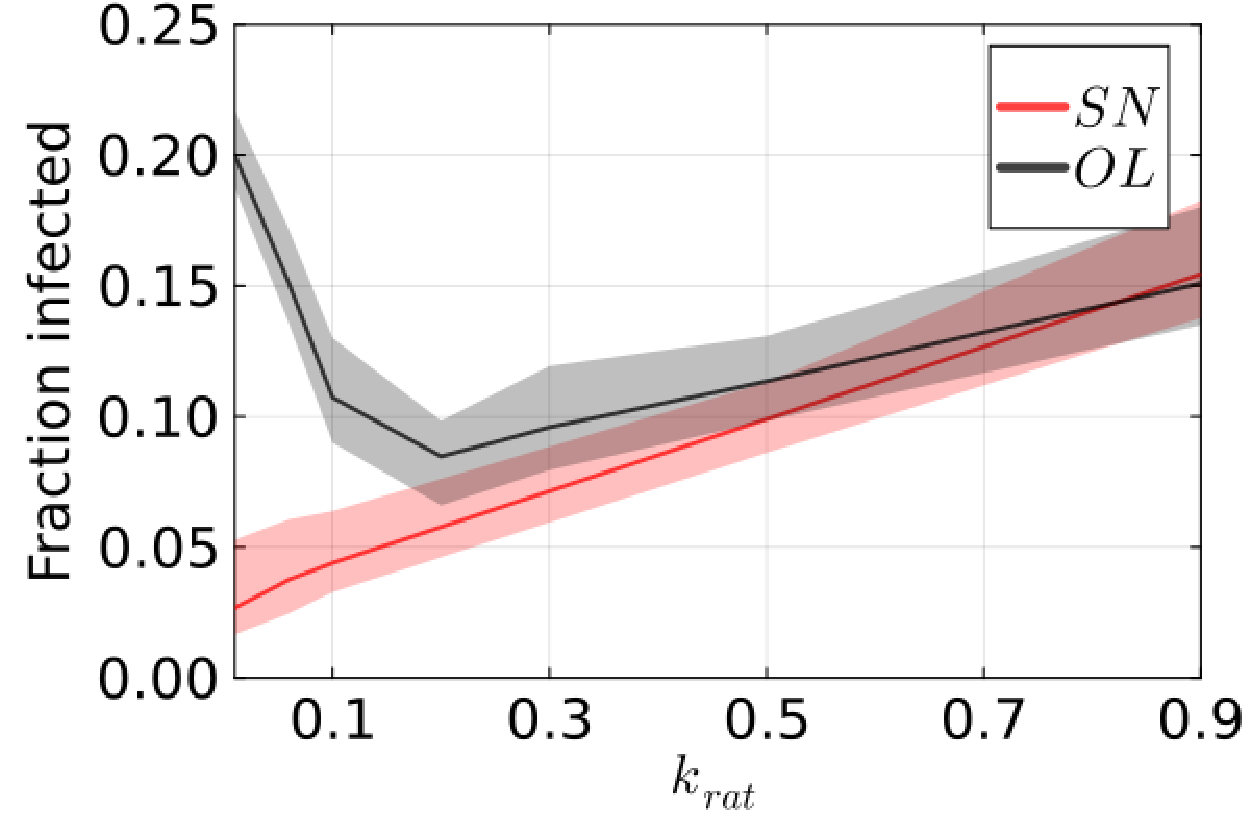}
\par\end{centering}
\caption{\label{fig:Fig3_1_2} \textbf{Fraction of infected individuals as a function of rationality.}  
When agents rely only on learning (OL), infection levels show a clear minimum at intermediate rationality, reflecting an optimal trade-off between exploration and exploitation. Highly irrational agents (large $k_{\mathrm{rat}}$) fail to process information effectively, sustaining high infection rates, while highly rational agents (small $k_{\mathrm{rat}}$) explore too little and oscillate between free-riding and full compliance. Including social norms (SN) removes these oscillations, as normative feedback stabilizes decisions and generally reduces infections. Yet, for very irrational agents, excessive dependence on social cues can backfire—individual learning alone may perform better.
}
\end{figure}

Figure~\ref{fig:Fig3_1_2} illustrates how outbreak size depends on the rationality parameter $k_{\mathrm{rat}}$, which measures agents’ sensitivity to payoff differences. The relationship is distinctly non-monotonic: epidemic size reaches a minimum at intermediate rationality. When agents are highly irrational (large $k_{\mathrm{rat}}$), they cannot effectively discriminate between payoffs and vaccinate inconsistently, sustaining high infection levels. When agents are overly rational (very small $k_{\mathrm{rat}}$), they rarely explore alternatives, leading to oscillations between free-riding and full compliance and producing recurrent outbreaks.  

Including social-norm consultation stabilizes these dynamics. By integrating normative feedback from neighbors, individuals rely less on strict payoff optimization, reducing the pathological cycles seen under perfect rationality. Overall, bounded rationality enhances collective welfare: moderate stochasticity promotes exploration, coordination, and stable cooperation, in line with previous experimental and computational results \cite{2024Charalambous}.

\subsection{Regret}

\begin{figure}[h!]
\centering
\includegraphics[scale=0.325]{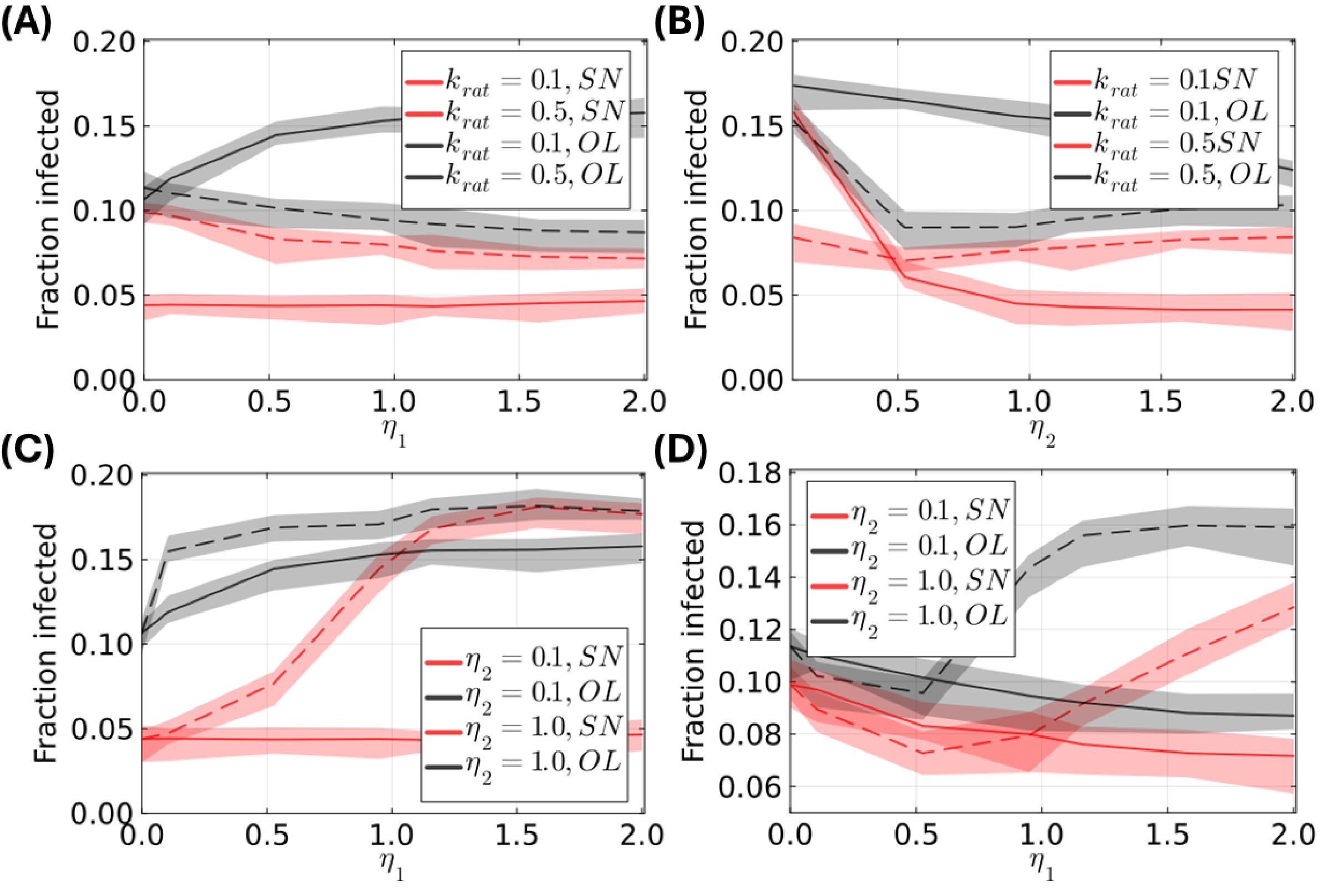}
%
%

\caption{\label{fig:Fig3_2_1}
\textbf{Behavioral impact of regret and greediness parameters.}
\textbf{(A)} Fraction of infected individuals vs.\ regret strength $\eta_{1}$.
With greediness fixed at $\eta_{2}=1.0$, rational agents ($k_{\mathrm{rat}}=0.1$) perform worse as regret increases without norms, while outcomes remain stable when norms are included. For less rational agents ($k_{\mathrm{rat}}=0.5$), higher $\eta_{1}$ improves results in both cases: regret fosters adaptive exploration among boundedly rational individuals but triggers emotional overreaction in highly rational ones.
\textbf{(B)} Fraction of infected individuals vs.\ greediness $\eta_{2}$.
At fixed regret strength $\eta_{1}=1.0$, reducing greediness benefits rational agents in both norm and no-norm scenarios, with larger gains under normative influence. For less rational agents, the effect is non-monotonic: extreme greed (low $\eta_{2}$) or indifference (high $\eta_{2}$) both worsen outcomes, whereas intermediate values yield lower infection levels. Moderate greed thus enhances exploration and learning stability.
\textbf{(C)} Fraction of infected individuals vs.\ regret strength $\eta_{1}$ for $k_{\mathrm{rat}}=0.1$.
\textbf{(D)} Same for $k_{\mathrm{rat}}=0.5$.
For rational, greedy agents ($k_{\mathrm{rat}}=0.1$, $\eta_{2}=0.1$), increasing $\eta_{1}$ erases the distinction between norm and no-norm conditions, indicating that individuals cease to learn from peers. For less rational agents ($k_{\mathrm{rat}}=0.5$), an intermediate $\eta_{1}$ minimizes infections, showing that moderate regret promotes behavioral correction and coordination.}
\end{figure}

Figures~\ref{fig:Fig3_2_1}A and~\ref{fig:Fig3_2_1}B show how the regret parameters—the strength of regret $\eta_{1}$ and the degree of greediness $\eta_{2}$—affect epidemic outcomes. In Fig.~\ref{fig:Fig3_2_1}A, the influence of regret varies with agents’ rationality $k_{\mathrm{rat}}$. For highly rational agents ($k_{\mathrm{rat}}=0.1$), increasing $\eta_{1}$ worsens results without norms, as excessive sensitivity induces emotional overreaction and unstable learning. With social norms present, this instability disappears because normative feedback dampens overcorrection. For moderately rational agents ($k_{\mathrm{rat}}=0.5$), intermediate regret improves performance in both settings, reducing infections through adaptive exploration. The right panel of Fig.~\ref{fig:Fig3_2_1}B further shows that lower greediness (higher $\eta_{2}$) enhances collective welfare, especially under normative influence, suggesting that prosocial restraint complements regret-based adaptation. Overall, regret operates as a double-edged mechanism: moderate levels promote coordination and stability, whereas excessive regret destabilizes decision-making.

Figures ~\ref{fig:Fig3_2_1}C and ~\ref{fig:Fig3_2_1}D maps infection levels across the $\eta_{1}$–$\eta_{2}$ space. For highly rational and greedy agents ($k_{\mathrm{rat}}=0.1$, $\eta_{2}=0.1$), stronger regret cancels the benefit of norms, as individuals become emotionally rigid and cease learning effectively. For less rational agents ($k_{\mathrm{rat}}=0.5$), an intermediate $\eta_{1}$ minimizes infections, showing that regret-driven feedback can stabilize collective behavior when cognitive precision is limited. This non-monotonic pattern mirrors empirical evidence that moderate anticipated regret fosters preventive action, while excessive regret or greed weakens cooperation.

These dynamics align with longitudinal findings on influenza vaccination: individuals who initially avoided vaccination but later experienced regret were more likely to vaccinate the following year, whereas those who regretted vaccinating were less likely to repeat it \cite{2006Chapman}. The model reproduces this asymmetry—moderate regret promotes adaptive correction, while excessive regret in payoff-driven agents induces overreaction. Consistent with regret theory \cite{2008Hayashi}, strong regret aversion can lead to counterproductive behavior. Thus, regret acts as both corrective and fragile—adaptive when moderate, distortionary when extreme—linking affective learning, bounded rationality, and social norms.

%

\subsection{Uncertainty}

\begin{figure}[h!]

\begin{centering}
\includegraphics[scale=0.32]{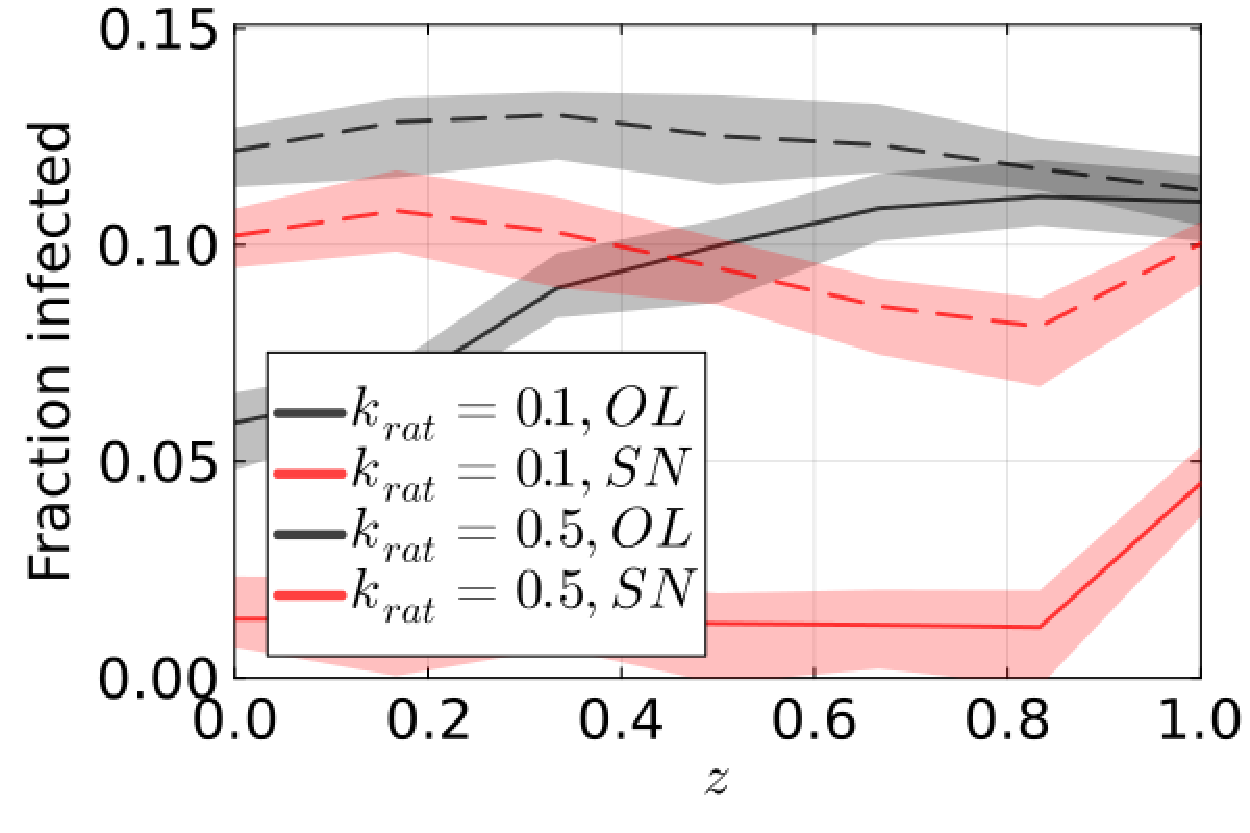}
\par\end{centering}
\caption{\label{fig:Fig3_3} \textbf{Fraction of infected individuals vs.\ percentage of observed neighbors $z$.} For rational agents ($k_{\mathrm{rat}}=0.1$) relying only on individual learning, reduced information about neighbors’ states promotes caution—raising vaccination and reducing outbreaks. In contrast, for less rational agents ($k_{\mathrm{rat}}=0.5$), limited information hampers learning and increases infections. When social norms are included, rational agents again achieve lower infection levels than when acting independently. For less rational agents, moderate information scarcity can be beneficial by preventing overreaction, but excessive uncertainty ultimately worsens outcomes.}
\end{figure}

Figure~\ref{fig:Fig3_3} illustrates how informational uncertainty, expressed as the number of observed neighbors $n_{\mathrm{Obs}}$, shapes epidemic outcomes under different levels of rationality. For rational agents ($k_{\mathrm{rat}}=0.1$) relying solely on personal experience, reduced information (lower $n_{\mathrm{Obs}}$) induces precautionary behavior and greater vaccination, resulting in smaller outbreaks. In contrast, less rational agents ($k_{\mathrm{rat}}=0.5$) fare worse under limited information: uncertainty magnifies decision noise and raises infection levels.  

Including social norms alters this relationship. Rational agents achieve substantially lower infection rates across all $n_{\mathrm{Obs}}$, as normative feedback compensates for missing epidemiological information. For less rational agents, moderate uncertainty can even be advantageous—curbing overreliance on unreliable cues—though excessive opacity once again disrupts coordination. Overall, the effect of uncertainty is non-monotonic: partial ignorance can foster caution and cooperation when supported by reasoning or social guidance, but too much uncertainty undermines collective stability.

\subsection{Interventions}

We examine how external interventions targeting different normative dimensions affect epidemic outcomes. Each intervention has strength $\gamma = 1.0$, driving the targeted norm toward a desired value $G = 0.5$, below the level reached without intervention (see Fig.~\ref{fig:Fig3}B). Figure~\ref{fig:Fig3_4}A shows the resulting infection fraction as a function of the infectivity rate $\beta$.  

Effects vary across normative domains. Interventions on personal norms $y$ produce the largest reduction in infections, indicating that self-endorsed moral standards are both flexible and strongly linked to behavior. Targeting injunctive expectations $\tilde{y}$ yields moderate gains, while interventions on empirical expectations $\tilde{x}$—reflecting perceived peer behavior—have the weakest impact, as descriptive norms adjust only through observed collective behavior and resist external influence. 

These results are in agreement with empirical work suggesting that personal norms are both highly predictive of behavior and relatively malleable through targeted interventions \cite{2021Catola}. 
Furthermore, they also align with experimental evidence that injunctive norms are more adaptable than descriptive ones \cite{2024Geber,2022Woike,2021Szekely}. Overall, interventions are most effective when shaping internal moral evaluations rather than perceptions of others’ actions, emphasizing the importance of belief formation in public health communication.

\subsubsection{Regret}

\begin{figure}[h!]
\centering
\includegraphics[scale=0.215]{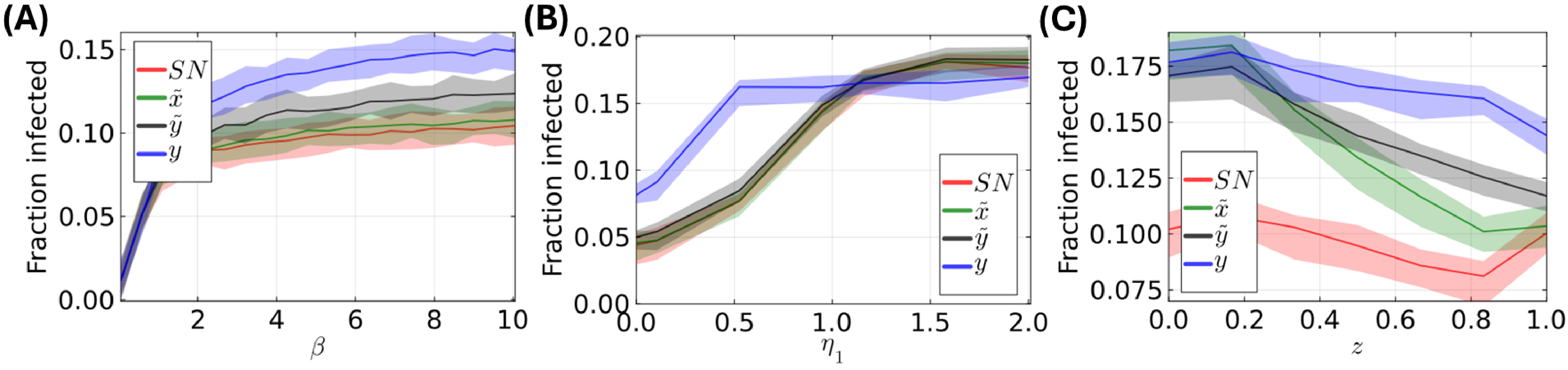}

\caption{\label{fig:Fig3_4}
\textbf{Intervention effects through different normative channels.}
\textbf{(A)} Fraction of infected individuals vs.\ infectivity rate $\beta$ ($G=0.5$, $\gamma=1.0$).
Interventions through personal norms $y$ are most effective, confirming their greater malleability compared with social norms. Empirical expectations $\tilde{x}$ respond least, reflecting their resistance to external influence.
\textbf{(B)} Fraction of infected individuals vs.\ regret strength $\eta_{1}$ under interventions ($G=0.5$, $\gamma=1.0$).
For rational agents ($k_{\mathrm{rat}}=0.1$), effectiveness shifts with rising $\eta_{1}$—initially strongest via personal norms $y$, then increasingly through social norms $\tilde{x}$ and $\tilde{y}$. As regret sensitivity grows, collective expectations gain leverage over personal standards.
\textbf{(C)} Fraction of infected individuals vs.\ percentage of observed neighbors $z$ under interventions ($G=0.5$, $\gamma=1.0$).
For moderately rational agents ($k_{\mathrm{rat}}=0.5$), interventions are most effective through personal norms $y$ and least through empirical expectations $\tilde{x}$ when information is complete. Under high uncertainty (low $z$), this hierarchy reverses, with social norms becoming the main influence channels.}
\end{figure}

Figure~\ref{fig:Fig3_4}B illustrates how interventions modulating regret strength $\eta_{1}$ affect epidemic outcomes across normative targets. Each intervention has intensity $\gamma = 1.0$ and target value $G = 0.5$. For rational agents ($k_{\mathrm{rat}} = 0.1$), a crossover in effectiveness emerges: at low intervention levels, targeting personal norms $y$ yields the strongest infection reduction, whereas stronger interventions on social norms—particularly empirical ($\tilde{x}$) and injunctive ($\tilde{y}$) expectations—become more effective.  

This shift reflects how cognitively precise agents internalize social feedback. As reasoning accuracy increases, altering collective expectations better promotes coordinated vaccination, while individual-focused interventions lose influence once social coupling dominates. Overall, the success of regret-based strategies depends jointly on emotional sensitivity $\eta_{1}$ and the normative dimension targeted. Leveraging collective regret signals can thus align individual heuristics with collective welfare under bounded rationality.

%

\subsubsection{Uncertainty}

Figure~\ref{fig:Fig3_4}C illustrates how intervention effectiveness varies with informational uncertainty, measured by the percentage of observed neighbors $z$, across different normative targets. For moderately rational agents ($k_{\mathrm{rat}} = 0.5$), a clear reversal emerges as information becomes scarce. With full visibility, interventions on personal norms $y$ are most effective, since well-informed individuals rely primarily on internal moral evaluation. As uncertainty increases (lower $z$), social norms gain influence, with injunctive expectations $\tilde{y}$ and normative expectations $\tilde{x}$ eventually surpassing personal norms in effectiveness. 

This transition shows that under limited visibility, individuals depend more on social inference than on self-guided judgment. Hence, uncertainty shifts the optimal intervention channel from personal to social norms, revealing that the informational environment critically determines behavioral leverage.  

Across parameter sweeps, three mechanisms consistently sustain coordination: (i) intermediate rationality balances exploration and learning, (ii) moderate regret promotes adaptive correction, and (iii) uncertainty redistributes normative influence, transferring effectiveness from self-oriented to socially mediated pathways.

%

\section{Conclusions}

This study introduces a behavioral epidemic model that integrates regret, uncertainty, and social norms within a unified agent-based framework. Drawing on psychology, behavioral economics, and communication research, it formalizes how anticipated regret, fear, and normative expectations interact with bounded rationality to shape vaccination behavior under collective risk. A key conceptual advance is the reconciliation of two forms of uncertainty: in behavioral epidemiology, uncertainty reflects limited confidence in risk judgments, whereas in epidemic game theory it represents incomplete information about others’ states. Here, structural uncertainty (imperfect network knowledge) generates psychological uncertainty (variable confidence and trust), both shaping probabilistic choices. By combining the Loomes–Sugden regret–rejoice mechanism with uncertainty-dependent trust and dynamic norms, the model provides a psychologically grounded account of preventive decision-making that links individual cognition to collective outcomes.

Simulations reveal several robust patterns.  
First, intermediate rationality optimizes collective outcomes: when decisions are neither random nor perfectly calculated, agents coordinate more effectively, consistent with prior work on stochastic choice and cooperation \cite{1998Fudenberg,2016Bicchieri}.  
Second, regret plays a dual role. Moderate regret promotes adaptive vaccination, while excessive regret or greed destabilizes behavior, echoing evidence that emotional salience can both motivate and distort preventive action \cite{2006Chapman,2016Brewer,2008Hayashi,2015Bleichrodt}.  
Third, uncertainty shows a similar non-linear effect: limited information encourages precautionary vaccination, but excessive ambiguity undermines coordination \cite{2020Hubner,2025Zhuang}.  
Finally, social norms mitigate informational gaps by providing cues that compensate for missing epidemiological knowledge, consistent with evidence that descriptive norms gain influence under uncertainty \cite{2015Gelfand,2016Tankard}.

The model also clarifies how the impact of norm-based interventions depends on information availability. When information is abundant, targeting personal norms is most effective, as moral self-consistency drives compliance \cite{2016Bicchieri,2019Yamin}. Under uncertainty, injunctive and descriptive norms dominate as individuals rely more on peer approval and observation. These shifts align with experimental evidence that informational context determines whether internal or social motives sustain cooperation \cite{2023Moehring,2016Nyborg}. The greater flexibility of injunctive norms relative to descriptive ones \cite{2024Geber,2022Woike,2021Szekely} highlights the importance of moral communication and institutional signaling when behavioral feedback is limited.

Methodologically, the Regret–Uncertainty model illustrates how agent-based modeling can bridge psychological realism and epidemiological dynamics. By embedding cognitive, emotional, and normative heterogeneity into decision rules, it extends previous vaccination ABMs \cite{2014Oraby,2012dOnofrio,2021Wu,2024Charalambous} and generalizes experience-weighted attraction learning to include affective feedback and endogenous norm evolution. Unlike imitation models that treat norms as external, this framework internalizes moral and descriptive expectations, enabling targeted interventions and linking behavioral economics with computational epidemiology.

In sum, the model unites bounded rationality, emotion, and social influence within a single formalism that reproduces key behavioral regularities—memory effects, optimal rationality, emotional modulation, and uncertainty-driven conformity. Beyond vaccination, it advances the study of norm change and collective behavior \cite{2016Bicchieri,2022Andrighetto}, showing how emotions and trust dynamically reweight moral and pragmatic motives. Future work could explore coevolving networks, where interaction structures and norms adapt jointly \cite{2021Gavrilets,2016Friedkin}, and cross-cultural variation in norm sensitivity, as tight and loose cultures differ in conformity and sanctioning \cite{2011Gelfand}. Combining behavioral experiments and online field data with simulations will help test and refine these mechanisms in real-world contexts \cite{2022Andrighetto,2019Yamin}.

\section*{Acknowledgments}
\vspace{-1em}
C.C. has received funding from the European Union’s Horizon 2020 research and innovation program under the Marie Sklodowska-Curie Grant Agreement No 101034403.
\vspace{-1.5em}
\section*{Data Availability Statement}
\vspace{-1em}
This is a computational study and the only data considered is the result of simulations of the coordination game.
To reproduce the research one should recreate the simulations. We facilitate this by describing in detail the
model and tools used in the main text. The datasets generated during the current study are available from the
corresponding author on reasonable request. Replication code can be found in GitHub: \\
\verb+https://github.com/Christos3788/Regret-Uncertainty-and-Bounded-Rationality-in-norm-driven-decisions+.

\vspace{-1.5em}
\bibliography{BiblioBehEpi} 


\end{document}